# A critical examination of compound stability predictions from machine-learned formation energies

Christopher J. Bartel[1*], Amalie Trewartha[1], Qi Wang[2], Alexander Dunn[1,2], Anubhav Jain[2], Gerbrand Ceder[1,3*]

[1]Department of Materials Science & Engineering, University of California, Berkeley, Berkeley, CA 94720, USA

[2]Energy Technologies Area, Lawrence Berkeley National Laboratory, Berkeley, CA 94720, USA

[3]Materials Sciences Division, Lawrence Berkeley National Laboratory, Berkeley, CA 94720, USA

[*]correspondence to cbartel@berkeley.edu, gceder@berkeley.edu

## Abstract

Machine learning has emerged as a novel tool for the efficient prediction of materials properties, and claims have been made that machine-learned models for the formation energy of compounds can approach the accuracy of Density Functional Theory (DFT). The models tested in this work include five recently published compositional models, a baseline model using stoichiometry alone, and a structural model. By testing seven machine learning models for formation energy on stability predictions using the Materials Project database of DFT calculations for 85,014 unique chemical compositions, we show that while formation energies can indeed be predicted well, all compositional models perform poorly on predicting the stability of compounds, making them considerably less useful than DFT for the discovery and design of new solids. Most critically, in sparse chemical spaces where few stoichiometries have stable compounds, only the structural model is capable of efficiently detecting which materials are stable. The non-incremental improvement of structural models compared with compositional models is noteworthy and encourages the use of structural models for materials discovery, with the constraint that for any new composition, the ground-state structure is not known *a priori*. This work demonstrates that accurate predictions of formation energy do not imply accurate predictions of stability, emphasizing the importance of assessing model performance on stability predictions, for which we provide a set of publicly available tests.

## Introduction

Machine learning (ML) is emerging as a novel tool for rapid prediction of material properties.[1–6] In general, these predictions are made by fitting statistical models on a large number of data points. Because of the scarcity of well-curated experimental data in materials science, this input data is often obtained from Density Functional Theory (DFT) calculations housed in one of the many open materials databases.[7–12] In principle, once these models are trained on this immense set of quantum chemical data, the determination of properties for new materials can be made in orders-of-magnitude less time using the trained models compared to computationally expensive DFT calculations.

Of particular interest is the use of machine learning to discover new materials. The combinatorics of materials discovery make for an immensely challenging problem – if we consider the possible combinations of just four elements ($A$, $B$, $C$, $D$), from any of the ~80 elements that are technologically relevant, there are already ~1.6 million quaternary chemical spaces to consider. This is before we consider such factors as stoichiometry ($ABCD_2$, $AB_2C_3D_4$, etc.) or crystal structure, each of which add substantially to the combinatorial complexity. The Inorganic Crystal Structure Database (ICSD) of known solid-state materials contains ~$10^5$ entries,[13] several orders of magnitude less than the $10^{10}$ quaternary compositions identified as plausible using electronegativity- and charge-based rules.[14] This suggests that 1) there is ample opportunity for new materials discovery and 2) the problem of finding stable materials may resemble the needle-in-a-haystack problem, with many unstable compositions for each stable one. The immensity of this problem is a natural fit for high-throughput machine learning techniques.

In this work, we closely examine whether recently published machine learning models for formation energy are capable of distinguishing the relative stability of chemically similar materials and provide a roadmap for doing the same for future models. We show that although the formation energy of compounds from elements can be learned with high accuracy using a variety of machine learning approaches, these learned formation energies do not reproduce DFT-calculated relative stabilities. While the accuracy of these models for formation energy approaches the DFT error (relative to experiment), DFT predictions benefit from a systematic cancellation of error[15,16] when making stability predictions, while ML models do not. Of particular concern for most ML models is the high rate of materials predicted to be stable that are not stable by DFT, impeding the use of these models to efficiently discover new materials. As a result, we propose more critical evaluation methods for machine learning of thermodynamic quantities.

## Results and Discussion

### The relationship between formation energy and stability

A necessary condition for a material to be used for any application is stability (under some conditions). The thermodynamic stability of a material is defined by its Gibbs energy of decomposition, $\Delta G_d$, which is the Gibbs formation energy, $\Delta G_f$, of the specified material relative to all other compounds in the relevant chemical space. Temperature-dependent thermodynamics are not yet tractable with high-throughput DFT and have only sparsely been addressed with ML,[17] so material stability is primarily assessed using the decomposition enthalpy, $\Delta H_d$, which is approximated as the total energy difference between a given compound and competing compounds in a given chemical space.[15,16,18,19] For the purpose of this study we will directly compare ML predictions and DFT calculations of $\Delta H_d$, hence the lack of entropy contributions is not an issue.

The quantity $\Delta H_d$ is obtained by a convex hull construction in formation enthalpy ($\Delta H_f$)-composition space. **Figure 1a** shows an example for a binary *A*-*B* space, having three known compounds, $A_4B$, $A_2B$, and $AB_3$. The convex hull is the lower convex enthalpy envelope which lies below all points in the composition space (blue line). Stable compositions lie on the convex hull, and unstable compositions lie above the hull. $A_4B$ is *unstable* (above the hull), so $\Delta H_d > 0$ and is calculated as the distance in $\Delta H_f$ between $A_4B$ and the convex hull of stable points. $AB_3$ is stable (on the hull), so $\Delta H_d < 0$ and is calculated as the distance in $\Delta H_f$ between $AB_3$ and a hypothetical convex hull constructed without $AB_3$ (dashed line). $|\Delta H_d|$ is therefore the minimum amount that $\Delta H_f$ must decrease for an unstable compound to become thermodynamically stable or the maximum amount that $\Delta H_f$ can increase for a stable compound to remain stable. We used $\Delta H_d$ in this work instead of the more common, "energy above the hull", because the former provides a distribution of values for stable compounds whereas the latter is equal to 0 for all stable compounds. The convex hull construction, described here for a binary system, generalizes for chemical spaces comprised of any number of elements.

Hence, while $\Delta H_f$ quantifies to what extent a compound may form from its elements, the thermodynamic quantity that controls phase stability is $\Delta H_d$ and arises from the competition between $\Delta H_f$ for all compounds within a chemical space. As we show later, while formation enthalpies can be of the order of several eV the value of $\Delta H_d$ is typically 1-2 orders of magnitude smaller. In addition, thermodynamic stability is highly nonlinear in $\Delta H_d$ around zero, as negative values indicate stable compounds, whereas positive values are unstable or metastable compounds. Although $\Delta H_d$ determines stability, the standard thermodynamic property that is predicted by ML models is the absolute $\Delta H_f$.[20–29] This is in large part because $\Delta H_f$ is intrinsic to a given compound, whereas $\Delta H_d$ inherently depends upon a compound's stability relative to neighboring compositions, making $\Delta H_d$ less robust to learn directly.

Using data available in the Materials Project (MP),[16] we applied the convex hull construction to obtain $\Delta H_d$ for 85,014 inorganic crystalline solids (the majority of which are in the ICSD) and compare $\Delta H_d$ to $\Delta H_f$ in **Figure 1b**. It is clear that effectively no linear correlation exists between $\Delta H_d$ and $\Delta H_f$, except for the trivial case where only a single compound exists in a chemical space ($\Delta H_d = \Delta H_f$), which is true for only ~3% of materials in MP. While $\Delta H_f$ somewhat uniformly spans a wide range of energies (mean ± average absolute deviation = -1.42 ± 0.95 eV/atom), $\Delta H_d$ spans much smaller energies (0.06 ± 0.12 eV/atom), suggesting $\Delta H_d$ is the more sensitive or subtle quantity to predict (histograms of $\Delta H_f$ and $\Delta H_d$ are provided in **Figure 1b**). Still, while no linear correlation exists between $\Delta H_d$ and $\Delta H_f$, and $\Delta H_d$ occurs over a much smaller energy range, it would be possible for $\Delta H_f$ models to predict $\Delta H_d$ as long as the relative differences in $\Delta H_f$ within a given chemical space are predicted with accuracy comparable to the range of variation in $\Delta H_d$ or if they would benefit from substantial error cancellation when comparing the energies of compounds with similar chemistry.

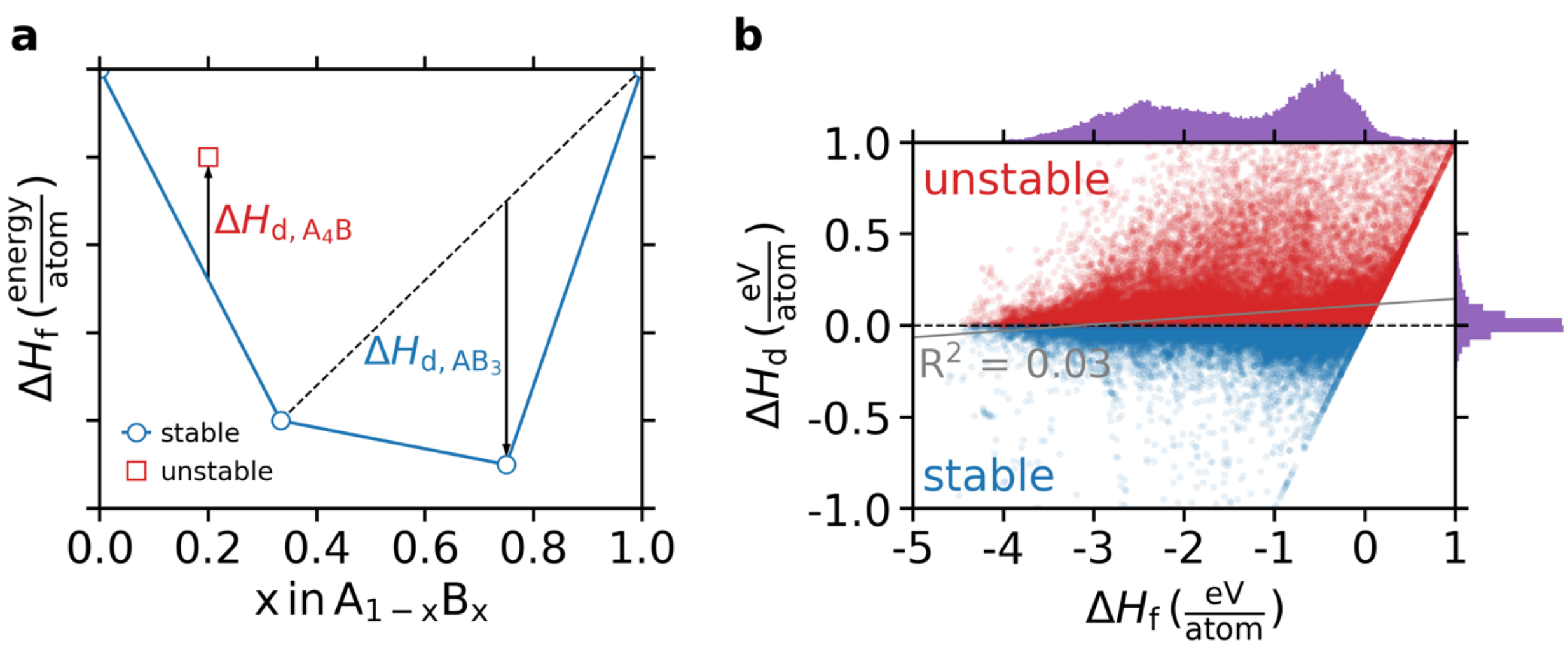

**Figure 1. a)** Illustration of the convex hull construction to obtain the decomposition enthalpy, $\Delta H_d$, from the formation enthalpy, $\Delta H_f$. **b)** The decomposition enthalpy, $\Delta H_d$, shown against the formation enthalpy, $\Delta H_f$, for 85,014 ground-state entries in Materials Project, indicating effectively no correlation between the two quantities. The strong linear correlation that is present at $\Delta H_d = \Delta H_f$ arises for chemical spaces that contain only one compound (~3% of compounds). These compounds were excluded from the correlation coefficient, $R^2$, determination. A normalized histogram of $\Delta H_f$ ($\Delta H_d$) is shown above (along the right side of) panel **(b)**. Both histograms are binned every 10 meV/atom.

## Learning formation energy from chemical composition

Machine learning material properties requires that an arbitrary material is "represented" by a set of attributes (features). This representation can be as simple as a vector corresponding to the fractional amount of each element in the compound (e.g., $Li_2O$ = [0, 0, 2/3, 0, 0, 0, 0, 1/3, 0, 0, …], where the length of the vector is the number of elements in the periodic table), or a vector that includes substantial physical or chemical information about the material. In the search for new materials, the structure is rarely known *a priori*, and instead a list of compositions with unknown structure is screened for stability, i.e., the possibility that a thermodynamically stable structure exists at that composition. In this case, the material representation is constructed only from the chemical formula without including properties such as the geometric (i.e., lattice and basis) or electronic structure. These models, which take as input the chemical formula and output thermodynamic predictions, are henceforth referred to as *compositional models* here.

In this work, we assessed the potential for five recently introduced compositional representations – *Meredig*,[20] *Magpie*,[21] *AutoMat*,[22] *ElemNet*,[23] and *Roost*[24] – to predict the stability of compounds in MP. *Meredig*, *Magpie*, and *AutoMat* include chemical information for each element in their material representations from quantities such as atomic electronegativities, radii, and elemental group. Each of these compositional representations were trained using gradient-boosted regression trees (XGBoost[30]). *ElemNet* and *Roost* differ in that no *a priori* information other than the stoichiometry is used as input. For *ElemNet*, a deep learning architecture maps the stoichiometry input into formation energy predictions. For *Roost*, the stoichiometric representation and fit are simultaneously learned using a graph neural network. In addition, we included a baseline representation for comparison, *ElFrac*, where the representation is simply the stoichiometric fraction of each element in the formula, trained using XGBoost[30]. Because compositional models necessarily make the same prediction for all structures having the same formula, all analysis in this work was performed using the lowest energy (ground-state) structure in MP for each available composition. Additional details on the training of each model and the MP dataset is available in the **Methods** section.

Parity plots comparing $\Delta H_f$ in MP ($\Delta H_{f,MP}$) to machine-learned $\Delta H_f$ ($\Delta H_{f,pred}$) for each model are shown in **Figure 2**. It is clear that each published representation substantially improves upon the baseline *ElFrac* model, decreasing the mean absolute error (MAE) by 27-74%. This increased accuracy is attributed to the increased complexity of the representation. For most models, the MAE between MP and these ML models is comparable to the expected numerical disagreement between MP and experimentally obtained $\Delta H_f$,[8,16,31–33] implying a substantial amount of the information required to determine $\Delta H_f$ is contained in the composition (and not the structure). The success of ML models for predicting $\Delta H_f$ is not surprising considering the historical context of simple heuristics that perform relatively well at predicting the driving force for the formation of compounds from elements – e.g., the Miedema model.[34]

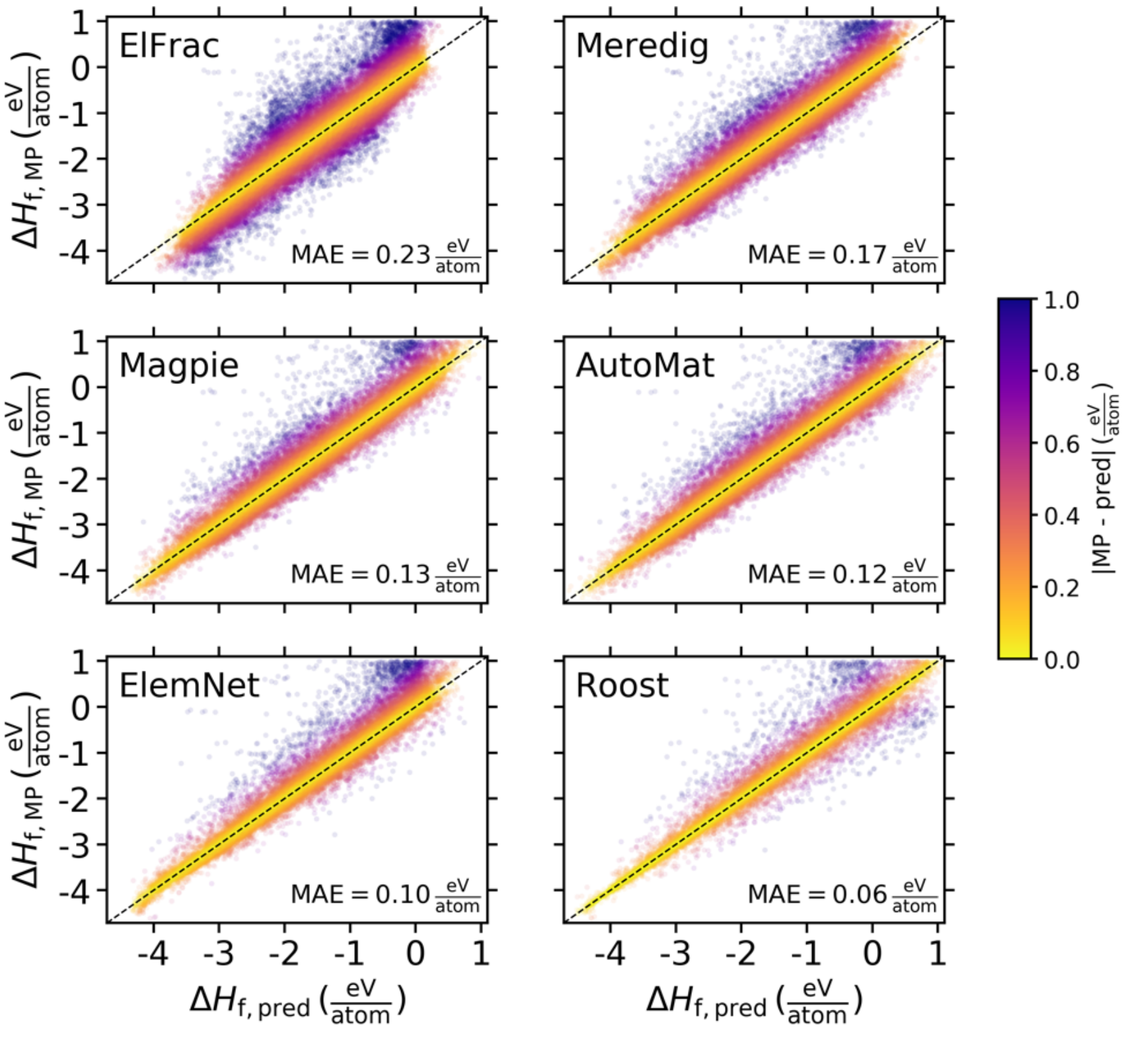

**Figure 2.** Parity plot for formation enthalpy predictions using six different machine learning models that take as input the chemical formula and output $\Delta H_f$. *ElFrac* refers to a baseline representation that parametrizes each formula only by the stoichiometric coefficient of each element. *Meredig*, *Magpie*, *AutoMat*, *ElemNet*, and *Roost* refer to the representations published in Refs. [20–24], respectively. $\Delta H_{f,pred}$ corresponds with ML predictions for aggregated hold-out sets during five-fold cross-validation of the Materials Project dataset (see **Methods** for details). $\Delta H_{f,MP}$ refers to the formation energy per atom in the MP database. The absolute error on $\Delta H_f$ is shown as the colorbar and the mean absolute error (MAE) is shown within each panel.

## Implicit stability predictions from learned formation enthalpies

While the mean absolute error (MAE) of the ML-predicted $\Delta H_f$ approaches the MAE between DFT and experiment for this quantity (~0.1-0.2 eV/atom)[8,16,31,32,35], the use of $\Delta H_f$ for stability predictions requires that the *relative* $\Delta H_f$ between chemically similar compounds is predicted more accurately. To assess the accuracy of the relative $\Delta H_f$, we reconstructed, for each ML model, the convex hulls for all chemical spaces using $\Delta H_{f,pred}$. Parity plots for $\Delta H_d$ are shown in **Figure 3**. Even though the quantity $\Delta H_d$ is on average much smaller than $\Delta H_f$, the MAE in predicting it is almost identical to the error in predicting $\Delta H_f$ (**Figure 2**), indicating very little error cancellation for the ML models when energy differences are taken in a chemical space, which is in contrast to the beneficial error cancellation for stability predictions with DFT.[15,16] In contrast to $\Delta H_f$, where all representations substantially improve the predictive accuracy from the baseline *ElFrac* model, for $\Delta H_d$, four of the five models (all except *Roost*) only negligibly improve upon the baseline model with MAE of ~0.10-0.14 eV/atom. Importantly, for the purposes of predicting stability, a difference of ~0.1 eV/atom can be the difference between a compound that is readily synthesizable and one that is unlikely to ever be realized.[36,37]

DFT calculations benefit from a systematic cancellation of errors that leads to much smaller errors for $\Delta H_d$ than for $\Delta H_f$, with MAE for $\Delta H_d$ as low as ~0.04 eV/atom for a substantial fraction of decomposition reactions.[16] Unfortunately, ML models do not similarly benefit from this cancellation of errors and instead appear to learn clusters in material space that have similar $\Delta H_f$, but they are generally unable to distinguish between stable and unstable compounds *within* a chemical space. It is notable that *Roost* substantially improves upon the other models. However, there are still strong signatures of inaccurate stability predictions in its parity plot (**Figure 3**), most notably in the ~vertical line at $\Delta H_{d,pred} = 0$ and ~horizontal line at $\Delta H_{d,MP} = 0$. These two lines indicate substantial disagreement between the actual and predicted stabilities for many compounds, despite the relatively low MAE.

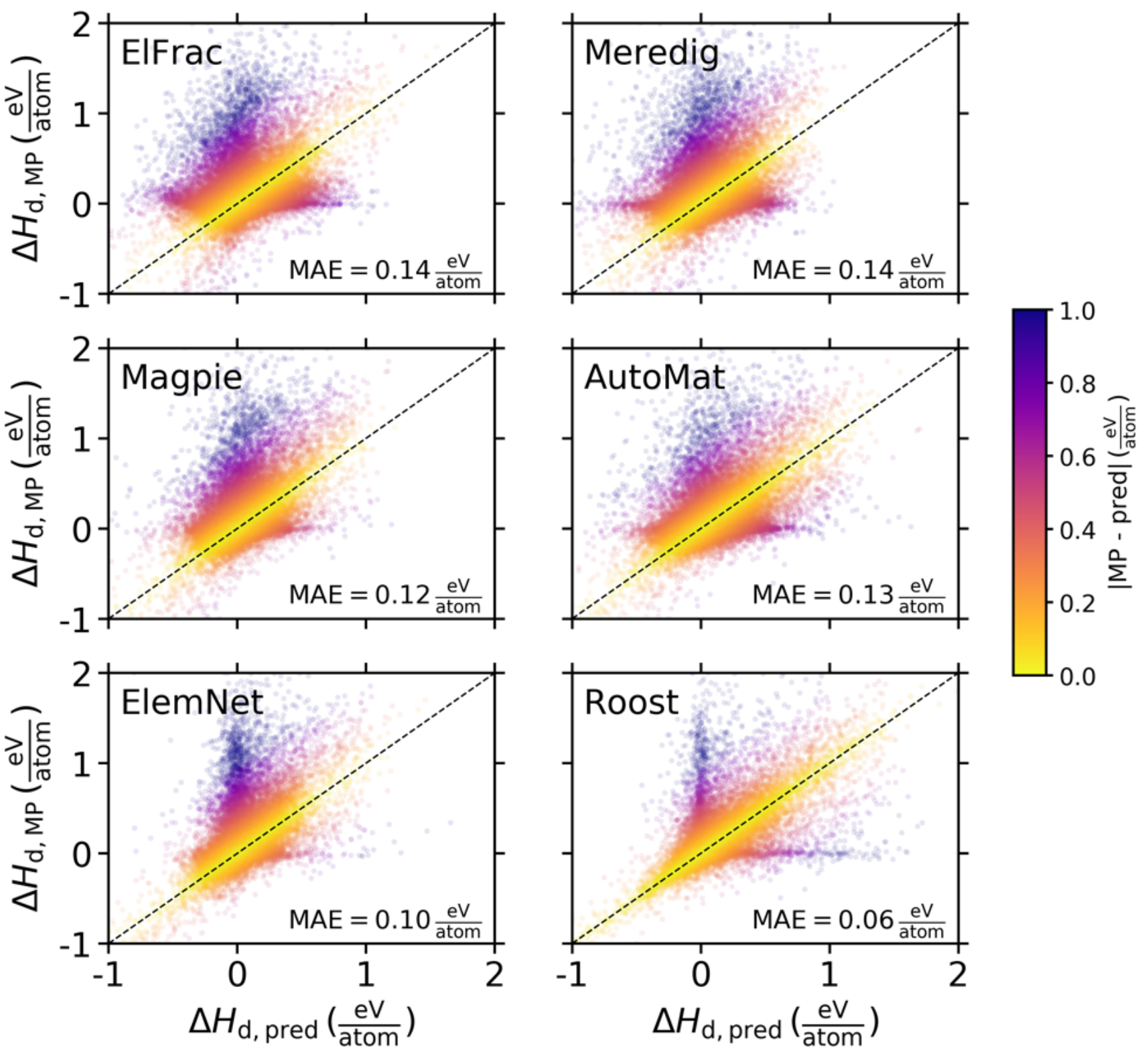

**Figure 3.** Parity plot for decomposition enthalpy predictions. $\Delta H_{\mathrm{d,MP}}$ results from convex hulls constructed with $\Delta H_{\mathrm{f,MP}}$ (**Fig. 2**). $\Delta H_{\mathrm{d,pred}}$ is obtained from convex hulls constructed with $\Delta H_{\mathrm{f,pred}}$ (**Fig. 2**). The annotations are the same as in **Fig. 2**.

The inability for compositional models to properly distinguish relative stability is further demonstrated by assessing how well the models classify compounds as stable (on the convex hull) or unstable (above the hull), as shown in **Figure 4**. 60% of the compounds in the MP dataset are not on the hull, so the classification accuracy of a naïve model that states that everything is unstable would be 60%. Five of the six models (all except *Roost*) only marginally improve upon this extremely naïve model accuracy (58-65%). Strikingly, *Roost* considerably outperforms the other compositional models (76% accuracy), despite using stoichiometry alone as input. Plausibly, this superior performance is due to the use of weighted soft attention mechanisms during training of the representation.[38] Although only the nominal chemical composition (element fraction) is used as input, the model learns a more meaningful representation of this input composition on a case-by-case basis during training. This is in contrast to the other compositional models, which have fixed stoichiometric representations and either include hand-picked elemental attributes such as electronegativity (*Meredig* and *Magpie*) or use deep learning (*ElemNet*). Notably, *AutoMat* uses a two-step process: first it rationally selects the most relevant elemental attributes from a large list using a decision tree model and then fits a regression model with the reduced feature space. Considering the modest classification accuracy by *AutoMat* (65%), despite the wide range of elemental attributes considered in its optimization, we speculate that further improvements in the clever selection of these attributes is unlikely to lead to transformative improvements in predicted stabilities. Instead, major improvements to compositional formation energy models will likely result from qualitative changes in model architecture, as in *Roost*, and not from optimizing the selection of elemental attributes.

While *Roost* improves considerably upon other compositional models, the accuracy, $F_1$ score, and false positive rate taken together do not inspire much confidence that any of these models can accurately predict the stability of solid-state materials (**Figure 4**). Of particular concern is the high false positive rates of 25-38%. This metric provides the likelihood that a compound predicted to be stable will not actually be. Further aggravating this situation is that the false positive rate reported here for the models is greatly underestimated compared to the false positive rate that is expected for new materials discovery. The MP database is largely populated with known materials extracted from the ICSD, and this results in ~40% of the entries in MP being on the hull. The fraction of all conceivable hypothetical materials (from which new materials will be discovered) that are stable is likely several orders of magnitude lower than 40%. This necessitates that searches for new materials cover a huge number of possible compounds, and false positive rates in excess of 25% would lead to an enormous amount of predicted materials which are not stable, limiting the ability for these ML models to efficiently accelerate new materials discovery.

A key consideration when discussing the accuracy in classifying compounds as stable or unstable is the choice of threshold for stability, which we have chosen to be $\Delta H_d = 0$. In materials discovery or screening applications, compounds are often considered potentially synthesizable even for $\Delta H_d > 0$ to consider potential inaccuracies in the predicted stabilities and account for a range of accessible metastability.[36,37] To probe the effects of moving this threshold for stability to higher or lower values of $\Delta H_d$, we show the receiver operating characteristic (ROC) curves for each model in **Figure S1**. As the threshold for stability moves to larger positive values, increases in model accuracy are concomitant with an increase in the rate of false positives and a decrease in the confidence that the compounds predicted to be stable are actually accessible. Conversely, as the threshold decreases below zero, the accuracy and false positive rate decrease together as less and less compounds meet this stricter threshold for stability. Ultimately, the conclusions we draw from setting a stability threshold of $\Delta H_d = 0$ are not affected by alternative stability thresholds.

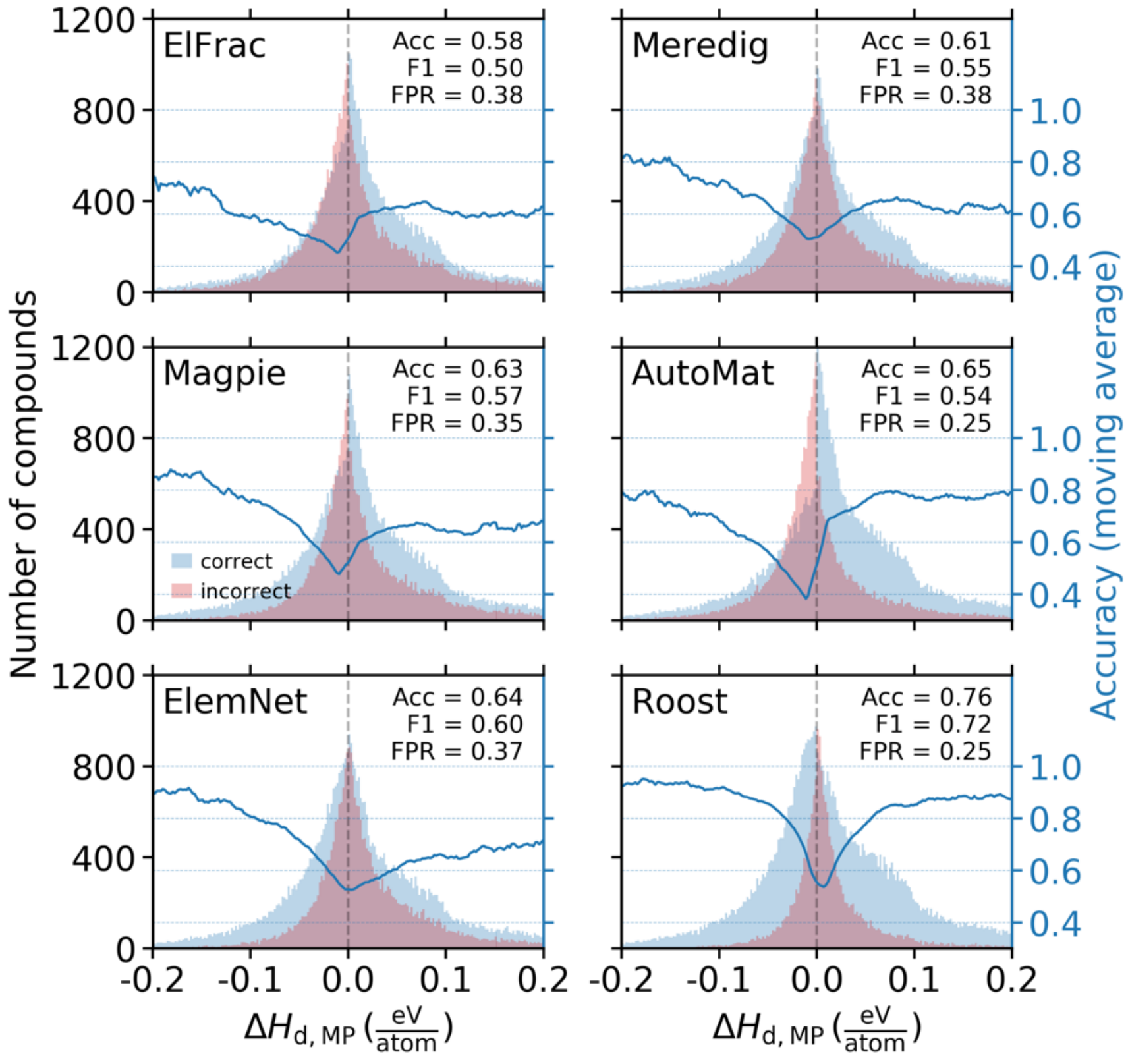

**Figure 4.** Classification of materials as stable ($\Delta H_d \leq 0$) or unstable ($\Delta H_d > 0$) using each of the six compositional models. "Correct" predictions are those for which the ML models and MP both predict a given material to be either stable or unstable. The histograms are binned every 5 meV/atom with respect to $\Delta H_{d,MP}$ to indicate how the correct and incorrect predictions and the number of compounds in our dataset vary as a function of the magnitude above or below the convex hull. Acc is the classification accuracy. F1 is the harmonic mean of precision and recall. FPR is the false positive rate. The moving average of the accuracy (computed within 20 meV/atom intervals) as a function of $\Delta H_{d,MP}$ is shown as a blue line (right axis). As expected, the accuracy is lowest near the chosen stability threshold of $\Delta H_{d,MP} = 0$.

## Predicting stability in sparse chemical spaces

While quantifying the accuracy of ML approaches on the entire MP dataset is instructive, it does not resemble the materials discovery problem because it assesses only the limited space of compositions that have been previously explored and therefore have many stable compounds. In order to simulate a realistic materials discovery problem, we identified a set of chemical spaces within the MP dataset that are sparse in terms of stable compounds. Lithium transition metal (TM) oxides are used as the cathode material for rechargeable Li-ion batteries and have attracted substantial attention for materials discovery in recent years. In particular, Li-Mn oxides have been considered as an alternative to $LiCoO_2$ utilizing less or no cobalt: e.g., spinel $LiMn_2O_4$,[39] layered $LiMnO_2$,[40] nickel-manganese-cobalt (NMC) cathodes,[41] and disordered rock salt cathodes.[42] For this work, the quaternary space, Li-Mn-TM-O with TM $\in$\{Ti, V, Cr, Fe, Co, Ni, Cu\}, is an attractive space to test the efficacy of these models, as it contains only 9 stable compounds and 258 more that are unstable in MP. We tested the potential for ML models to discover these stable compounds by excluding all 267 quaternary Li-Mn-TM-O compounds from the MP dataset and repeating the training of each model on $\Delta H_f$ with the remaining 84,747 compounds. We then applied each trained model to predict $\Delta H_f$ for the excluded Li-Mn-TM-O compounds and assessed their stability. Importantly, we are again concerned with DFT-calculated stability at 0 K, so we are not considering the potential for compounds in this quaternary space to be stabilized due to entropic effects (e.g., configurational disorder).

The $\Delta H_f$ parity plot for these 267 Li-Mn-TM-O compounds is shown in **Figure S2** and reveals that all models have a higher accuracy predicting $\Delta H_f$ for this subset of materials than for the entire dataset (**Figure 3**). The improved prediction of $\Delta H_f$ is likely because the compounds in this subset have strongly negative $\Delta H_f$ and are well-represented by the thousands of transition metal and lithium-containing oxides that comprise the MP dataset. Despite this improved accuracy on $\Delta H_f$, the models all have alarmingly poor performance in predicting $\Delta H_d$. In **Figure 5**, we show that none of the models are able to correctly detect more than three of the nine stable compounds, and even for the most successful model by this metric (*AutoMat*), the three true positives come with 24 false positives. It is noteworthy that in this experiment, the models are given a large head-start towards making these predictions because the composition space under investigation is restricted to those compounds that have DFT calculations tabulated in MP, which is biased towards stability compared to the space of all possible hypothetical compounds.

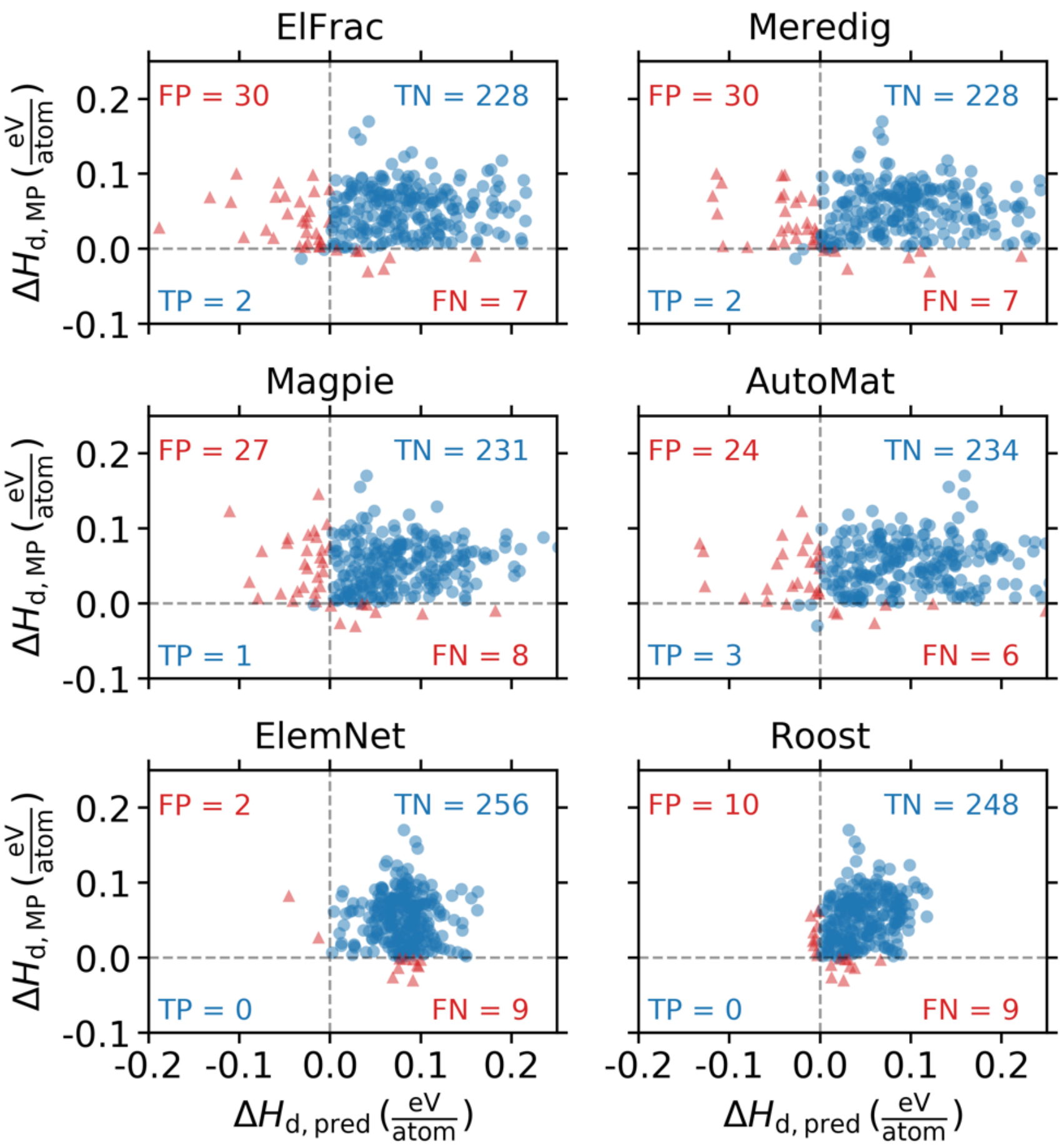

**Figure 5.** Re-training each model on all of MP minus 267 quaternary compounds in the Li-Mn-TM-O chemical space (TM ∈ {Ti, V, Cr, Fe, Co, Ni, Cu}) and obtaining $\Delta H_d$ using the predicted $\Delta H_f$ for each of the excluded compounds ($\Delta H_{d,pred}$) and comparing to stabilities available in MP, $\Delta H_{d,MP}$. FP = false positive, TP = true positive, TN = true negative, FN = false negative.

To account for the MP stability bias and more closely simulate a realistic materials discovery problem, we assessed the potential for these models to identify the nine stable MP compounds when considering a much larger composition space. Using the approach defined in Ref. [14], we produced 13,392 additional quaternary compounds in these seven Li-Mn-TM-O chemical spaces that obey simple electronegativity- and charge-based rules. For this expanded space of quaternary compounds, we used each compositional model (trained on all of MP minus the 267 Li-Mn-TM-O compounds) to predict $\Delta H_f$ and assessed their stability (**Table 1**). The compositional models each predict ~4-5% of these compounds to be stable, and all of the models fail to accurately predict the stability of more than two of the nine compounds that are actually stable in MP. A remarkable 139 compounds are predicted to be stable by all six models and 1,372 unique compounds are predicted to be stable by at least one model. While it is likely that the space of stable quaternary compounds in the Li-Mn-TM-O space has not yet been fully explored in MP (or by extension, the ICSD), our intuition suggests it is highly unlikely that the number of new stable materials in this well-studied space is orders of magnitude larger than the number of known stable materials. The false positive rates obtained on the entire MP dataset shown in **Figure 4** suggest ~25-38% of these predicted stable Li-Mn-TM-O compounds are not actually stable, and these rates are likely underestimated, as discussed previously. The magnitude of compounds predicted to be stable by the ML models, and their false positive rates, imply that these models will inevitably identify a large number of unstable materials as candidates for further analysis (either with DFT calculations or experimental synthesis). This substantially impedes the capability of these formation energy models to accelerate the discovery of novel compounds that can be synthesized.

**Table 1.** Predictions in the expanded Li-Mn-TM-O (TM ∈ {Ti, V, Cr, Fe, Co, Ni, Cu}) composition space. Candidate compounds were generated by combining all quaternary MP compounds in this space along with quaternary compounds generated by the approach described in Ref. [14], resulting in 13,659 candidates. Among these candidates, 9 compounds are calculated to be stable in MP. The stability of all candidates was assessed using each compositional model for $\Delta H_f$. Note that while all models correctly predict 1 of 9 MP-stable compounds to be stable, this compound is not the same for all models.

| | *ElFrac* | *Meredig* | *Magpie* | *AutoMat* | *ElemNet* | *Roost* |
|---|---|---|---|---|---|---|
| **candidate compounds** | 13,659 | 13,659 | 13,659 | 13,659 | 13,659 | 13,659 |
| **stable compounds in MP** | 9 | 9 | 9 | 9 | 9 | 9 |
| **compounds predicted stable** | 685 | 528 | 619 | 541 | 556 | 507 |
| **% predicted stable** | 5.0 | 3.9 | 4.5 | 4.0 | 4.1 | 3.7 |
| **pred. stable and stable in MP** | 1 | 1 | 1 | 1 | 2 | 1 |

## Direct training on decomposition energy

An alternative approach to consider is to train directly on $\Delta H_d$ instead of using ML-predicted $\Delta H_f$ to obtain $\Delta H_d$ through the convex hull construction. Note that direct training on $\Delta H_d$ is complicated by the fact that $\Delta H_d$ for a given compound is dependent upon $\Delta H_f$ for other compounds within a given chemical space. This is unlike $\Delta H_f$, which is intrinsic to a single compound. To assess the capability of each representation to directly predict stability, we repeated the analysis shown in **Figures 3-5** and **Table 1** but training on $\Delta H_d$. The performance of each model on the MP Li-Mn-TM-O dataset is shown in **Figure 6**, the performance on the expanded Li-Mn-TM-O space in **Table S1**, and results for $\Delta H_d$ on the entire MP dataset are shown in **Figures**

**S3-S4**. While the prediction accuracy (MAE and stability classification) on the entire MP dataset is typically comparable to or slightly better when training on $\Delta H_d$ (**Figures S3-S4**) instead of $\Delta H_f$ (**Figures 3-4**), the capability of the trained model to predict stability in sparse chemical spaces is even worse than when training on $\Delta H_f$ (**Figure 6**, **Table S1**).

None of the models are able to identify even one of the nine MP-stable quaternary compounds from the set of 267 Li-Mn-TM-O compounds in MP, and every model predicts all 267 Li-Mn-TM-O compounds to be unstable (**Figure 6**). It is especially notable that for all models except *Roost* and *ElemNet*, the predictions for all 267 quaternary compounds fall in a very small window (0.040 eV/atom < $\Delta H_{d,pred}$ < 0.082 eV/atom), suggesting the models only learn that all compounds in this space should be within the vicinity of the convex hull and do nothing to distinguish between chemically similar compounds. When the space of potential compounds is expanded to 13,659 compounds, only *Roost* and *ElemNet* predict any compound to be stable, but again, none of the nine MP stable compounds are predicted to be stable by any model (**Table S1**).

As an additional demonstration, all representations (except *Roost* – see Methods for details) were also trained as classifiers (instead of regressors), tasked with predicting whether a given compound is stable ($\Delta H_d \leq 0$) or unstable ($\Delta H_d > 0$). The accuracies, $F_1$ scores, and false positive rates are tabulated in **Table S2** and found to be only slightly better (accuracies < 80%, $F_1$ scores < 0.75, false positive rates > 0.15) than those obtained by training on $\Delta H_f$ (**Figure 4**) or $\Delta H_d$ (**Figure S4**).

Beyond the poor performance associated with these models, the direct prediction of $\Delta H_d$ (or classification of stable/unstable) is difficult to physically motivate because unlike $\Delta H_f$, $\Delta H_d$ is not an intrinsic property of a material but depends on the energy at other compositions with which it may be in competition. This non-locality of $\Delta H_d$ also depends on the completeness of a given phase diagram: as new materials are discovered in a chemical space, $\Delta H_d$ is subject to change for any compound in that space, even if that compound's energy itself does not change, complicating the application of ML models trained on $\Delta H_d$.

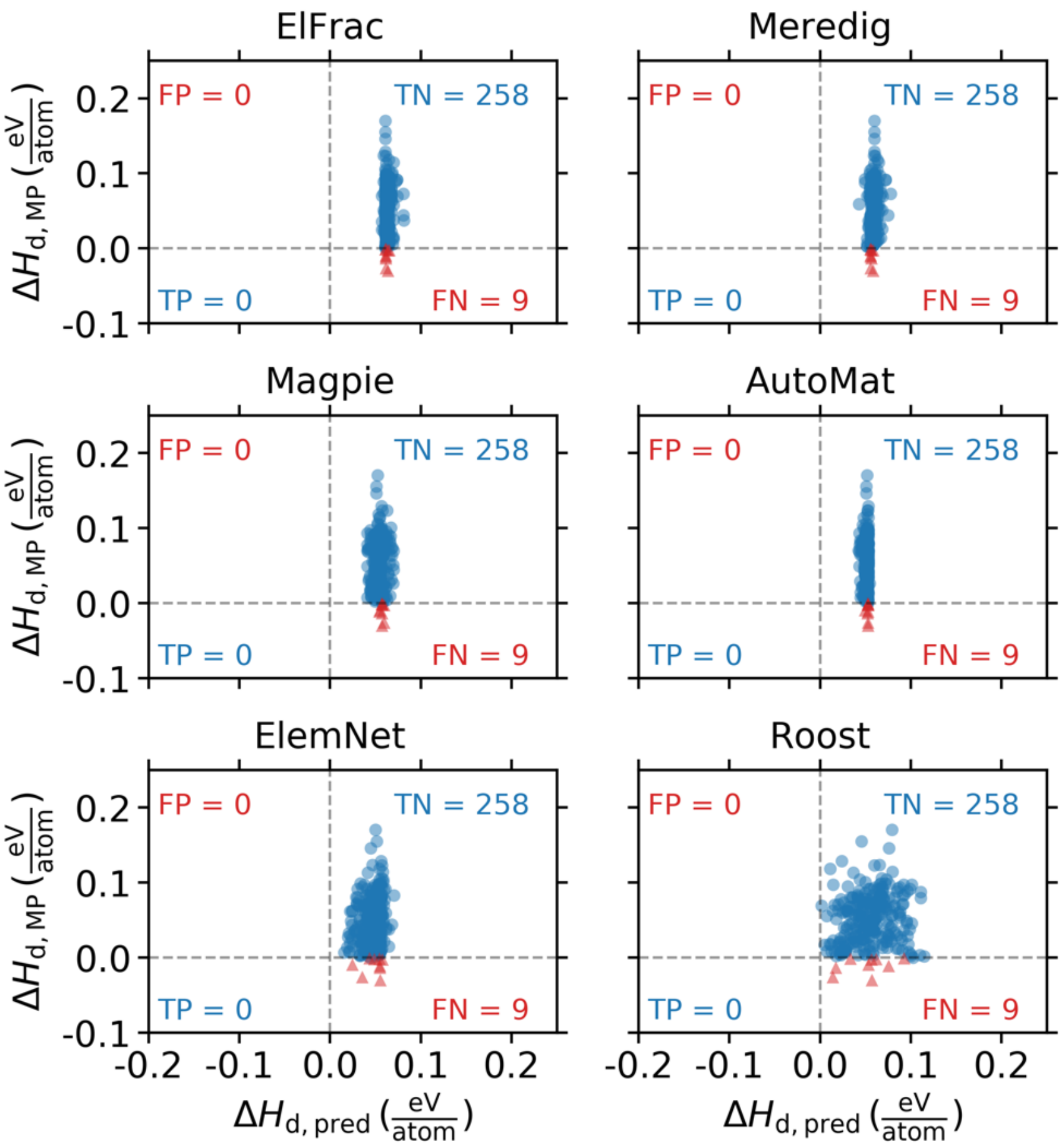

**Figure 6.** Re-training each model directly on $\Delta H_d$ on all of MP minus 267 quaternary compounds in the Li-Mn-TM-O chemical space (TM ∈ {Ti, V, Cr, Fe, Co, Ni, Cu}). Stability is determined by these direct predictions of $\Delta H_d$,pred and compared to stabilities available in MP, $\Delta H_{d,MP}$. FP = false positive, TP = true positive, TN = true negative, FN = false negative.

## Revisiting stability predictions with a structural representation

In addition to compositional models, representations that rely on the crystal structure for predicting formation energy have also received substantial attention in recent years.[25,27–29,43–45] These models perform a different task than compositional models because they evaluate the property of a material given both the composition and the structure. Nevertheless, it is interesting to assess whether these structural models can predict stability with improved accuracy relative to models that are given only composition.

Here we take the crystal graph convolutional neural network (*CGCNN*)[25] as a representative example of existing structural models. *CGCNN* is a flexible framework that uses message passing over the atoms and bonds of a crystal (see **Methods** for training details). In **Figure 7**, we show the performance of *CGCNN* on the same set of analyses as were shown for the compositional models in **Figures 2-5**: learning $\Delta H_f$ (**Figure 7a**), constructing convex hulls with those predicted $\Delta H_f$ to generate $\Delta H_d$ (**Figure 7b**), assessing the capability of these $\Delta H_{d,pred}$ values to classify materials as stable or unstable (**Figure 7c**), and probing the ability for this model to predict stability in the sparse Li-Mn-TM-O space (**Figure 7d**). It is clear that *CGCNN* improves substantially upon the direct prediction of $\Delta H_f$ (**Figure 7a**) and the implicit prediction of $\Delta H_d$ (**Figure 7b**), reducing the MAE by ~50% compared with the best performing compositional model (*Roost*). The extremely inaccurate predictions of $\Delta H_d$ near $\Delta H_{d,pred} = 0$ or $\Delta H_{d,MP} = 0$ that are observed in **Figure 3** for most compositional models, is also no longer present with *CGCNN* (**Figure 7b**). *CGCNN* displays an improved classification accuracy (80%) and a narrow distribution of incorrect stability predictions, only disagreeing with MP regarding the stability of compounds within the vicinity of $\Delta H_{d,MP} = 0$ (**Figure 7c**). Most impressively, *CGCNN* is relatively successful at finding the needles in the excluded Li-Mn-TM-O haystack, recovering five of the nine stable compounds with only six false positives (**Figure 7d**). In addition to the improved predictive accuracy, the parity plot for this excluded set looks fundamentally different than for the compositional models. In the compositional models (**Figure 5**), the parity plot is scattered, and there is effectively no linear correlation between the actual and predicted $\Delta H_d$, whereas for *CGCNN*, there is a strong linear correlation (**Figure 7d**).

The non-incremental improvement in stability predictions that arises from including structure in the representation is a strong endorsement for structural models and also sheds insight into the structural origins of material stability. While the thermodynamic driving force for forming a compound from its elements (formation energy) can be learned with high accuracy from only the composition, the structure dictates the subtle differences in thermodynamic driving force between chemically similar compounds and enables accurate machine learning predictions of material stability (decomposition energy). However, the obvious limitation of this approach is that it requires the structure as input, and the structure of new materials that are yet to be discovered is not known *a priori*. For example, because we do not know the ground-state structure for an arbitrary composition, we cannot repeat the test where we assess the ability of the ML model to find the stable Li-Mn-TM-O compounds among a large set of candidate compositions. Although *CGCNN* shows substantially improved performance in predicting material stability, these results are obtained using the DFT-optimized ground-state crystal structures as input.

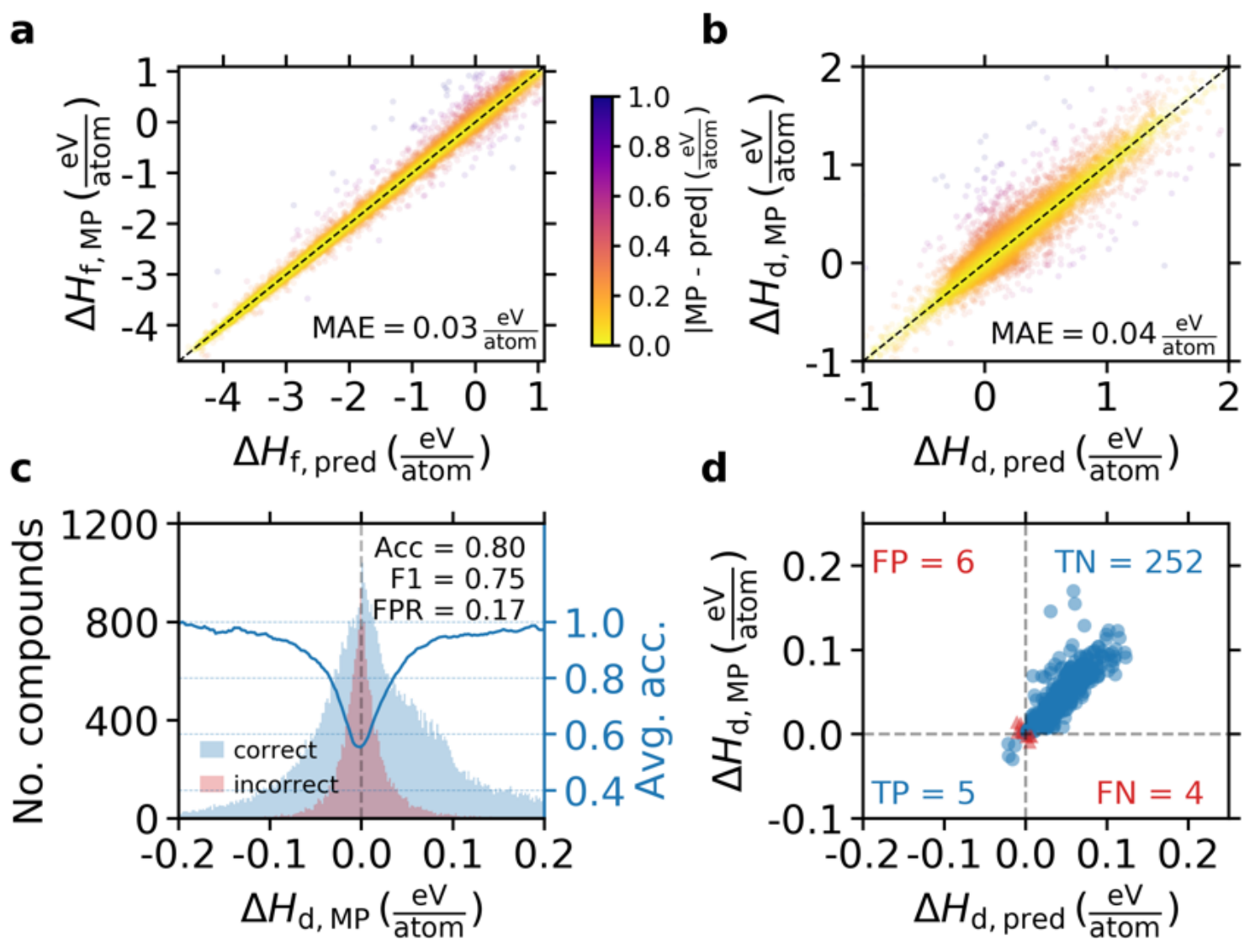

**Figure 7. a)** Repeating the analysis shown in **Figure 2** using *CGCNN*. Annotations are as in **Figure 2**. **b)** Repeating the analysis shown in **Figure 3** using *CGCNN*. Annotations are as in **Figure 3**. **c)** Repeating the analysis shown in **Figure 4** using *CGCNN*. Annotations are as in **Figure 4**. **d)** Repeating the analysis shown in **Figure 5** using *CGCNN*. Annotations are as in **Figure 5**.

## Quantifying error cancellation in ML models

While it is well known that DFT predictions of stability are enhanced because of systematic error cancellation,[15,16] it is not yet known if the errors made by ML formation energy models are completely random or if they too exhibit some beneficial cancellation of errors. In **Figure 8a**, we plot a hypothetical convex hull phase diagram in blue, labeled "ground truth". Next, we represent the effect of a systematic error in $\Delta H_f$ by shifting all points up in energy by the same amount ("systematic error", green). The relative stabilities of these systematically shifted points remain comparable to the ground truth despite the error in each $\Delta H_f$, illustrating beneficial cancellation of errors. Finally, we show the effect of a random $\Delta H_f$ error by shifting the ground truth $\Delta H_f$ by the same average magnitude as in the "systematic" case, but here in random amounts for each point (purple, "random error"). In this case, stabilities ($\Delta H_d$) for all points deviate substantially from the ground truth because there is little beneficial cancellation of $\Delta H_f$ errors.

The similar MAE on $\Delta H_f$ (**Figure 2**, **Figure 7a**) and $\Delta H_d$ (**Figure 3**, **Figure 7b**) for all models despite the much smaller range of energies spanned by $\Delta H_d$ (**Figure 1b**) make clear that the benefit of error cancellation is not fully realized for the ML models. Further, the set of tests on stability predictions discussed in this work (**Figure 4-6**, **Table 1**) show that the magnitude of error cancellation made by the compositional ML models remains insufficient to enable accurate stability predictions, especially in sparse chemical spaces. It is not clear, however, whether the improved prediction of stability by *CGCNN* arises from beneficial error cancellation within each chemical space or from decreasing the overall MAE from ~0.06 eV/atom (for the best performing compositional model – *Roost*) to ~0.03 eV/atom (for *CGCNN*).

To quantify the magnitude of error cancellation for the ML models, it is essential to establish a "random error" baseline for comparison. The random error baseline developed in this work utilizes random perturbations of the ground-truth $\Delta H_{f}$ ($\Delta H_{f,MP}$), where the perturbations were drawn from the discrete distribution of ML errors, $P[\Delta H_{f,MP} - \Delta H_{f,pred}]$, for each model. It follows that $\Delta H_{f,rand} = \Delta H_{f,MP} + P[\Delta H_{f,MP} - \Delta H_{f,pred}]$ ($\Delta H_{f,pred}$ is shown for each compositional model in **Figure 2** and for *CGCNN* in **Figure 7a**). With these randomly perturbed $\Delta H_{f,rand}$, we repeated the convex hull construction for all compounds in MP for comparison to the analysis of $\Delta H_{d}$ presented in **Figure 3** and stability classification presented in **Figure 4** (both of which rely on $\Delta H_{f,pred}$). In **Figure 8b**, the MAEs on $\Delta H_{d}$ and $F_1$ scores for classifying compounds as stable ($\Delta H_{d} \leq 0$) or unstable ($\Delta H_{d} > 0$) are compared for predictions based on $\Delta H_{f,pred}$ and $\Delta H_{f,rand}$ for all seven models. All models show higher MAE on $\Delta H_{d}$ and lower $F_1$ scores using $\Delta H_{f,rand}$ instead of $\Delta H_{f,pred}$ as input, demonstrating that the ML models do generally exhibit some degree of error cancellation.

The extent of error cancellation is model-dependent, and the worst-performing model in this work, *ElFrac*, exhibits the most substantial relative error cancellation, with 148% higher MAE on $\Delta H_{d}$ and 14% lower $F_1$ when using $\Delta H_{f,rand}$ instead of $\Delta H_{f,pred}$. For *ElFrac* (and to a lesser extent the other modestly performing compositional models), the high error cancellation likely arises because of the wide distribution of predicted $\Delta H_{f}$ ($P[\Delta H_{f,MP} - \Delta H_{f,pred}]$), which drives up the error of the random error baseline dramatically. *Roost* is remarkably shown to have considerable error cancellation (80% higher MAE on $\Delta H_{d}$ using $\Delta H_{f,rand}$) despite the MAE on $\Delta H_{d}$ using $\Delta H_{f,rand}$ already being competitive with the actual predictions ($\Delta H_{d}$ using $\Delta H_{f,pred}$) made by the other compositional models.

However, we emphasize that even benefiting from this substantial error cancellation, *Roost* is not able to detect the stability of compounds in sparse chemical spaces (as shown in **Figure 5**). The only model that performs suitably at this task is the structural representation, *CGCNN* (**Figure 7d**), and this model exhibits a much smaller degree of error cancellation (MAE on $\Delta H_{d}$ increased by 26% and $F_1$ decreased by 3% using $\Delta H_{f,rand}$). Because $\Delta H_{f,pred}$ for *CGCNN* is sufficiently accurate (MAE on $\Delta H_f$ = 34 meV/atom), the lack of error cancellation does not have a deleterious effect on stability predictions.

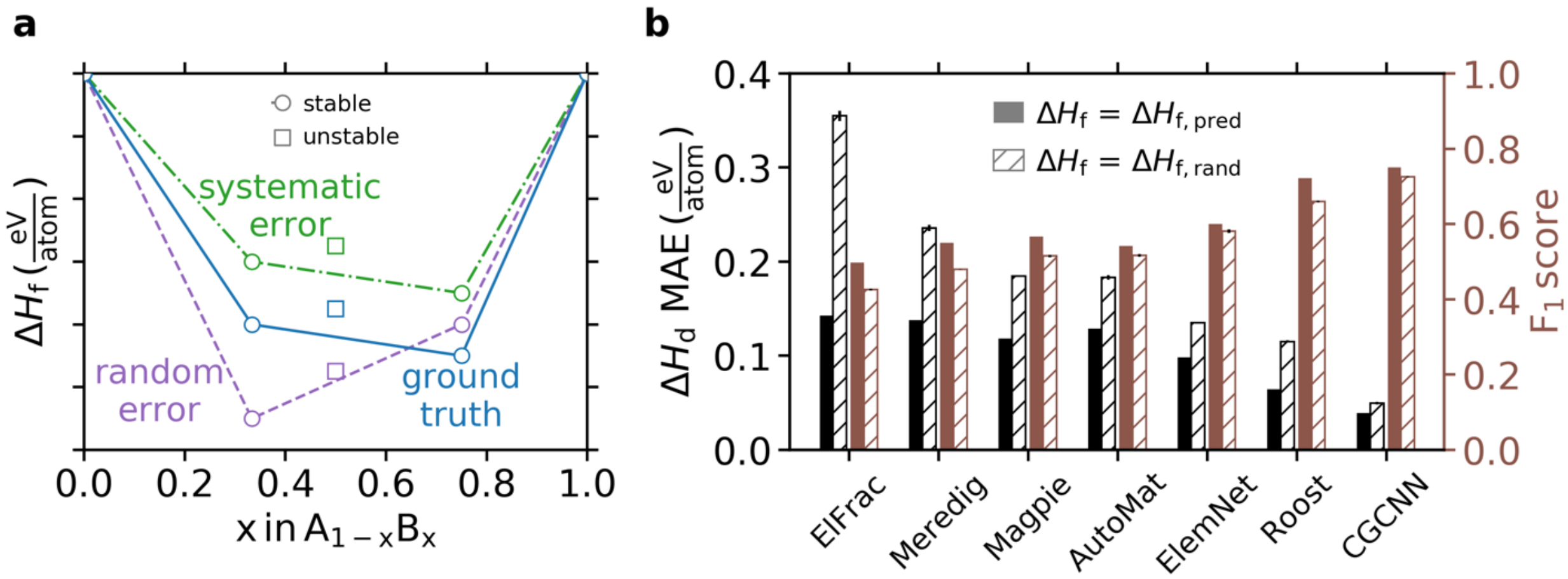

**Figure 8. a)** Schematic illustration contrasting how random and systematic errors on $\Delta H_f$ of the same average magnitude manifest as larger and smaller errors on predicted ground state lines ($\Delta H_d$). **b)** Comparing the performance on stability predictions using ML-predicted $\Delta H_f$ ($\Delta H_{f,pred}$, filled bars) and $\Delta H_f$ with perturbations drawn randomly from the distribution of $\Delta H_{f,pred}$ errors ($\Delta H_{f,rand} = \Delta H_{f,MP} + P[\Delta H_{f,MP} - \Delta H_{f,pred}]$, hatched bars). The mean absolute error (MAE) on $\Delta H_d$ is shown by the black bars (left axis). The $F_1$ score for classifying compounds as stable ($\Delta H_d \leq 0$) or unstable ($\Delta H_d > 0$) is shown by the brown bars (right axis). The results for the randomly perturbed case are averaged over three random samples with the standard deviation shown as an error bar. The standard deviation is too small to see in most cases.

## Outlook

There have been a number of recent successes in the application of machine learning for materials design problems. These models have given the impression that ML can predict formation energies with near-DFT accuracy.[20,23,46] However, the critical question of whether this implies that compound stability can be predicted by ML has not been rigorously assessed. In this work, we show that while indeed existing ML models can predict $\Delta H_f$ with relatively high accuracy from the chemical formula, they are insufficient to accurately distinguish stable from unstable compounds within an arbitrary chemical space. The error in predicting DFT-calculated $\Delta H_f$ by ML models is often compared favorably to the error DFT makes in predicting $\Delta H_f$ relative to experimentally obtained values. This comparison neglects the fact that the errors in DFT-calculated $\Delta H_f$ are beneficially systematic, whereas the errors made by the ML models are not. For DFT calculations, this leads to substantially lower errors for stability predictions ($\Delta H_d$) than for $\Delta H_f$. A similarly beneficial cancellation of errors does not occur for ML models and the errors in $\Delta H_d$ are comparable to $\Delta H_f$, inhibiting accurate predictions of material stability. Hence, while the claim that ML-predicted formation energies have similar errors as DFT compared to experiment is technically correct, it does not imply that in their current state ML models are as useful as DFT, or that ML can replace DFT for the computationally guided discovery of novel compounds. As new ML models for formation energy are developed, it is imperative to assess their viability as inputs for stability predictions and, most critically, for problems that resemble how the models would be implemented to address emerging materials design problems. In this work, we present a set of tests that facilitate this assessment and allow for direct comparison to existing ML models. All data and code required to repeat this set of stability analyses for the models shown in this work or any new model is available at https://github.com/CJBartel/TestStabilityML.

## Methods

### Materials Project data

All entries in the Materials Project (MP)[9] database were queried on July 26, 2019 using the Materials Project API.[47] This produced 85,014 unique non-elemental chemical formulas. For each chemical formula, we obtained the formation energy per atom, $\Delta H_f$, for all structures having that formula, and used the most negative (ground-state) $\Delta H_f$ for training the models and obtaining $\Delta H_d$ by the convex hull construction. MP applies a correction scheme to improve the agreement between DFT-calculated thermodynamic properties ($\Delta H_f$ and $\Delta H_d$) and experiment.[15,48,49] Additional details on the MP calculation procedure can be found at https://materialsproject.org/docs/calculations.

Although the MP database contains a wide range of inorganic crystalline solids, it is an evolving resource that periodically includes more and more compounds as they are discovered or calculated by the community. As such, the calculated $\Delta H_d$ that were used for training and testing each model are subject to change over time as new stable materials are added to the database. This fact is not unique to MP and is inherent in all open materials databases that would be considered for training and evaluating machine learning models on large datasets of DFT calculations. The $\Delta H_d$ and $\Delta H_f$ used for all compounds in this work are available within https://github.com/CJBartel/TestStabilityML.

**General training approach**

Five-fold cross validation was used to produce the model-predicted $\Delta H_{f,pred}$ shown in **Figure 2**. Each predicted value corresponds with the prediction made on that compound when it was in the validation set (i.e., not used for training). $\Delta H_{f,pred}$ was then used in the convex hull analysis to generate $\Delta H_{d,pred}$ shown in **Figure 3**, from which stability classifications were made as shown in **Figure 4**. For the Li-Mn-TM-O examples (**Figure 5** and **Table 1**), each model was trained on all MP entries except those 267 quaternary compounds belonging to the Li-Mn-TM-O chemical spaces. An analogous approach was used when training on $\Delta H_d$ instead of $\Delta H_f$ to generate the results shown in **Figure 6**, **Figure S3-S4**, **Table S1**, and **Table S2**.

**Compositional model training**

Three of the compositional representations—*ElFrac*, *Meredig*,[20] and *Magpie*[21]—were implemented using *matminer*[50] and trained using gradient boosting as implemented in XGBoost[30] with 200 trees and a maximum depth of 5. Preliminary tests showed XGBoost and these hyperparameters led to the highest accuracy of tested algorithms. *AutoMat*[22] was used as implemented in Ref. [22]. *Roost*[24] was trained for 500 epochs using an Adam optimizer with an initial learning rate of $5\times10^{-4}$ and an $L_1$ loss function. *ElemNet* was implemented as described in Ref. [23] using the Keras machine learning framework[51]. *ElemNet* was trained using an initial learning rate of $10^{-4}$ with an Adam optimizer for 200 epochs. 10% of the input data was set aside for validation, and the model weights from the epoch with best loss on the validation set were used for predictions.

Regarding training each representation as a classifier, for *ElFrac*, *Meredig*, *Magpie*, and *AutoMat*, we used an XGBoost classifier with the same parameters as used for regression. For *ElemNet*, we added a sigmoid activation function to the output and used cross entropy loss for training. All other aspects of these models were identical to those trained for regression. *Roost* was excluded from the classification analysis as modifying this representation to perform classification required more extensive changes than for the other representations.

Learning curves for all compositional models trained on $\Delta H_f$ are provided in **Figure S5** along with training and inference times in **Table S3**. The code used to train and evaluate all models is available at https://github.com/CJBartel/TestStabilityML.

***CGCNN* training**

We used a nested five-fold cross-validation to train the *CGCNN*[25] model for the MP $\Delta H_f$ dataset. As a general procedure for cross-validation, the dataset was split into five groups and each group was iteratively taken as a hold-out test set. For each fold, we split the training set to 75% training and 25% validation, thus the overall ratio of training, validation, and test was 60%, 20%, and 20%, respectively. The *CGCNN* model was iteratively updated by minimizing the loss (mean squared error, MSE) on the training set, and the validation score (mean absolute error, MAE) was monitored after each epoch. After 1000 epochs, the model with the best validation score was selected and then evaluated on the hold-out test set. Results of the five-fold hold-out test sets were accumulated as the final predictions of the dataset.

For the Li-Mn-TM-O case in which the test set is defined, we split the remaining compounds into five groups and iteratively took each group as the validation set (20%) and the remaining as the training set (80%). The best *CGCNN* model of each fold was selected as the one with the best validation score (MAE). We then applied the five *CGCNN* models to the 267 Li-Mn-TM-O test compounds and used the average of the predicted $\Delta H_f$ for each model.

## Acknowledgments

This work was primarily funded by the U.S. Department of Energy, Office of Science, Office of Basic Energy Sciences, Materials Sciences and Engineering Division under Contract No. DE-AC02-05-CH11231 (Materials Project program KC23MP). High Performance Computing resources were sponsored by the U.S. Department of Energy's Office of Energy Efficiency and Renewable Energy, located at NREL. This research also used the Savio computational cluster resource provided by the Berkeley Research Computing program at the University of California, Berkeley (supported by the UC Berkeley Chancellor, Vice Chancellor for Research, and Chief Information Officer) and the Lawrencium computational cluster resource provided by the IT Division at the Lawrence Berkeley National Laboratory (Supported by the Director, Office of Science, Office of Basic Energy Sciences, of the U.S. Department of Energy under Contract No. DE-AC02-05CH11231).

## Author contributions

CJB, AT, and GC conceived the project. CJB, AT, QW, and AD designed the project. AT, QW, and AD implemented the machine learning models. CJB performed the stability analysis, processed the results, and drafted the manuscript. AJ and GC supervised the project. All authors reviewed and edited the manuscript.

## Data availability

A public repository at https://github.com/CJBartel/TestStabilityML contains the following items: code for training each compositional model in this work, code for assessing the stability of compounds given predicted formation energies (for the models shown in this work or any new model), the formation and decomposition energy data for all models studied in this work, and code for generating all figures and tables in the main text and **Supplementary Information**.

## Competing interests

The authors declare no competing interests.

# Supplementary Information

# A critical examination of compound stability predictions from machine-learned formation energies

Christopher J. Bartel[1*], Amalie Trewartha[1], Qi Wang[2], Alexander Dunn[1,2], Anubhav Jain[2], Gerbrand Ceder[1,3*]

[1]Department of Materials Science & Engineering, University of California, Berkeley, Berkeley, CA 94720, USA

[2]Energy Technologies Area, Lawrence Berkeley National Laboratory, Berkeley, CA 94720, USA

[3]Materials Sciences Division, Lawrence Berkeley National Laboratory, Berkeley, CA 94720, USA

[*]correspondence to cbartel@berkeley.edu, gceder@berkeley.edu

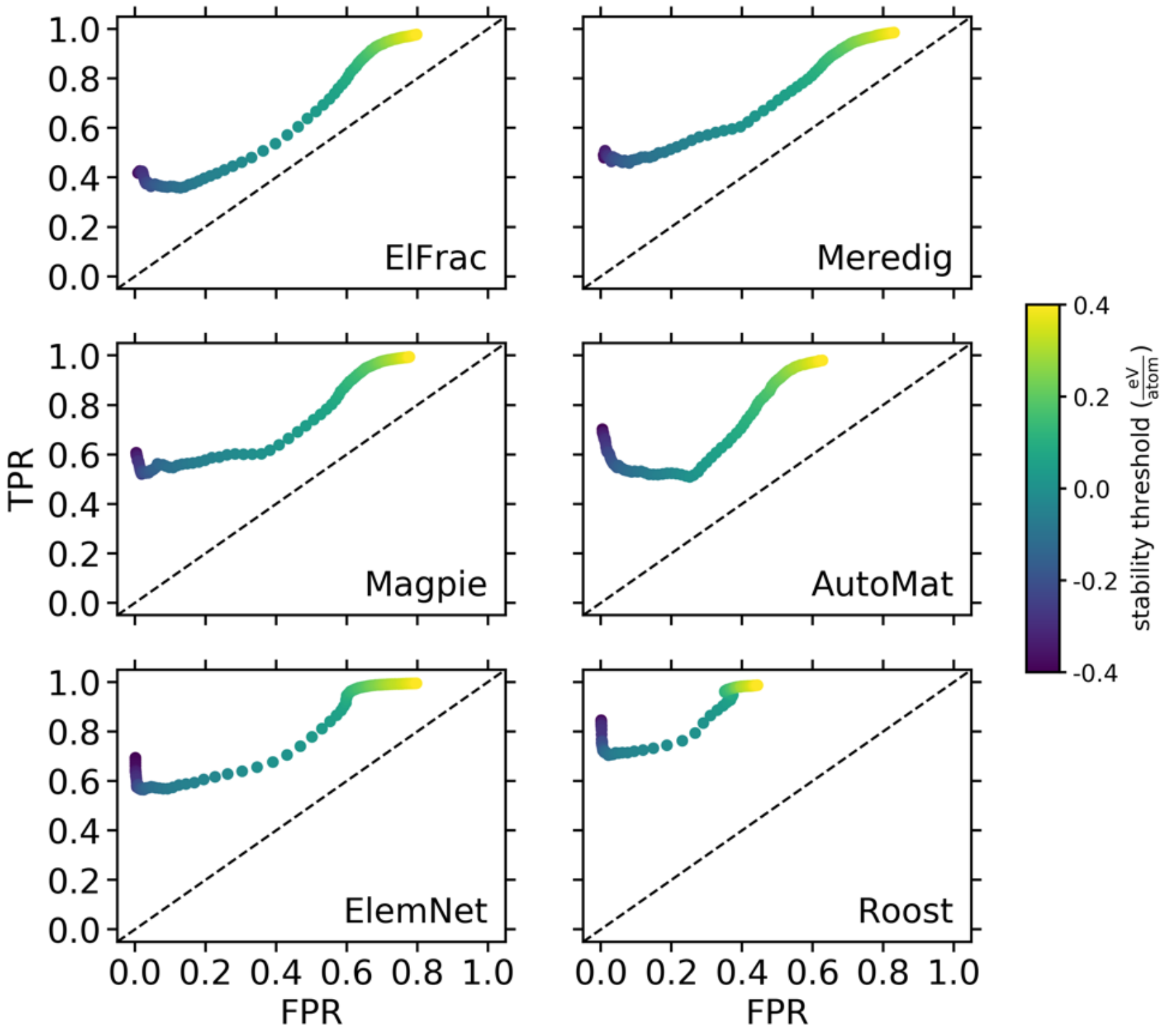

**Figure S1.** Receiver operating characteristic (ROC) curves for each model trained on $\Delta H_f$. TPR is the true positive rate and FPR the false positive rate. The colorbar indicates the stability threshold – i.e., a compound is classified as "stable" if $\Delta H_d$ is less than the stability threshold. Note that the models are trained on $\Delta H_f$ and are therefore insensitive to this changing threshold. Instead, the choice of threshold simply allows for an expanded analysis of the $\Delta H_f$ model performance on $\Delta H_d$ predictions.

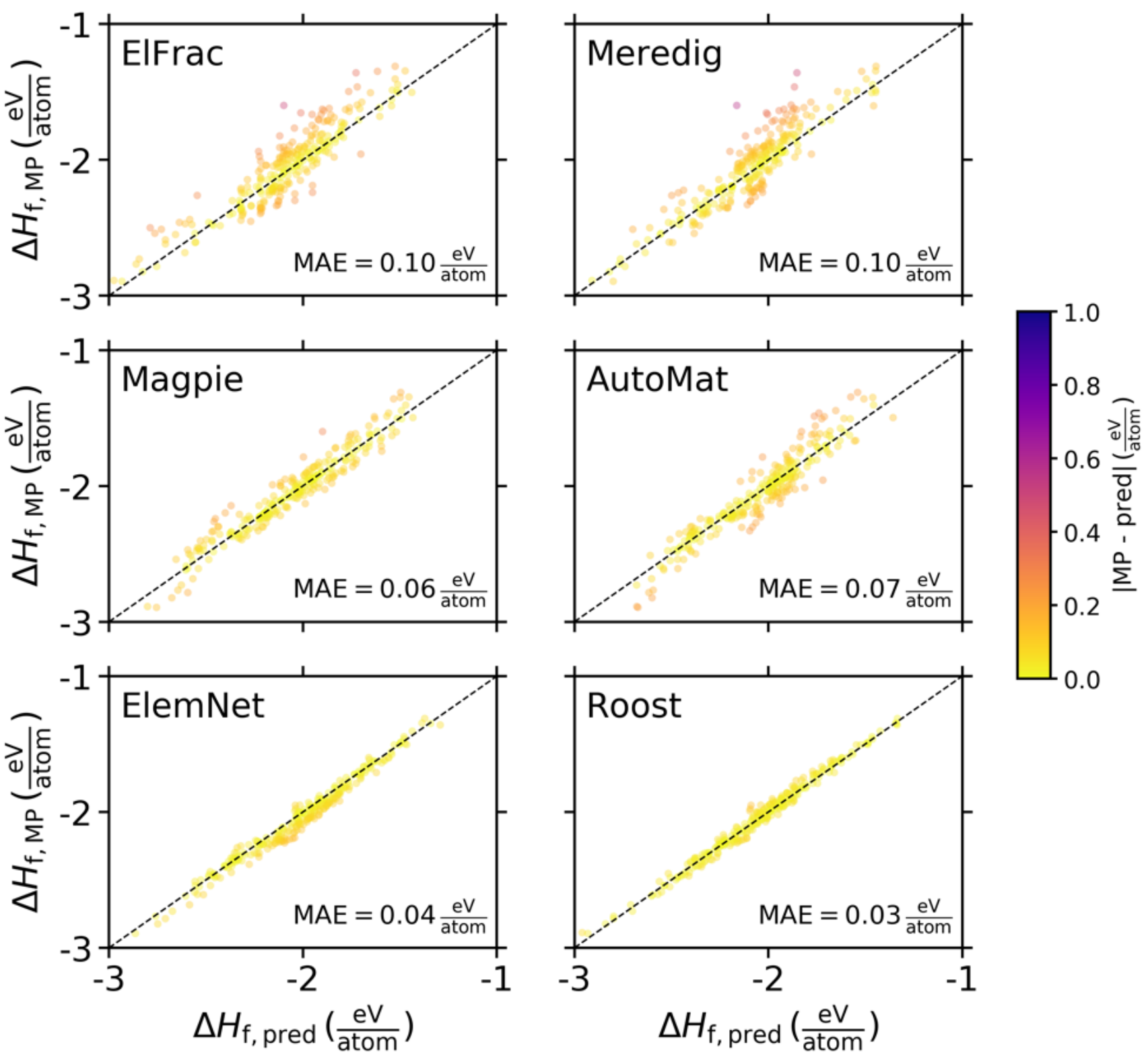

**Figure S2.** Re-training each model on all of MP minus 267 quaternary compounds in the Li-Mn-TM-O chemical space (TM ∈ {Ti, V, Cr, Fe, Co, Ni, Cu}) and predicting $\Delta H_f$ for each of the excluded compounds ($\Delta H_{f,pred}$) and comparing to MP, $\Delta H_{f,MP}$. All annotations are the same as in **Figure 2**.

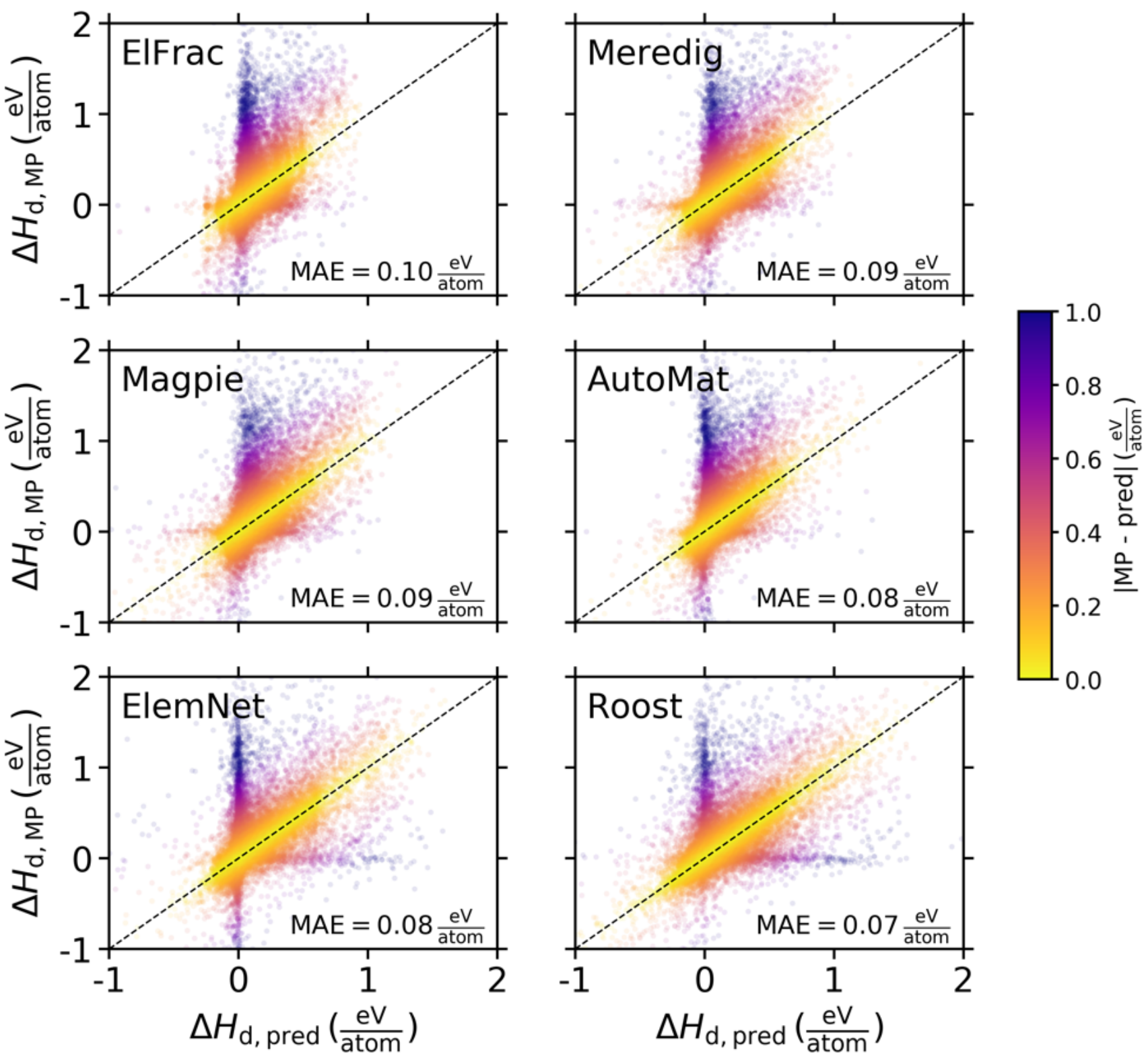

**Figure S3.** Reproducing **Figure 3** but training on $\Delta H_d$ instead of $\Delta H_f$. All annotations are the same as in **Figure 3**.

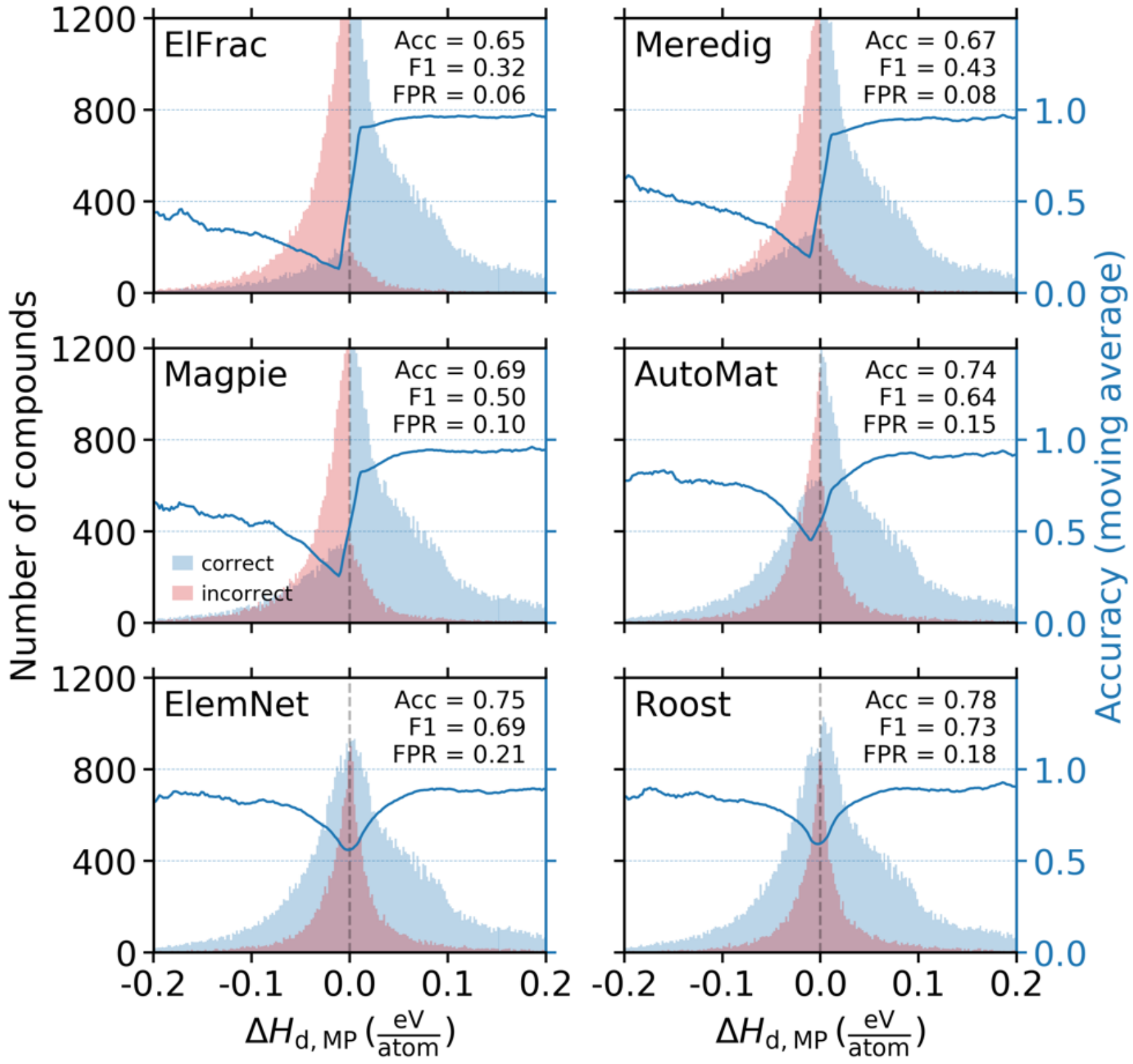

**Figure S4.** Reproducing **Figure 4**, but training on $\Delta H_d$ instead of $\Delta H_f$. All annotations are the same as in **Figure 4**.

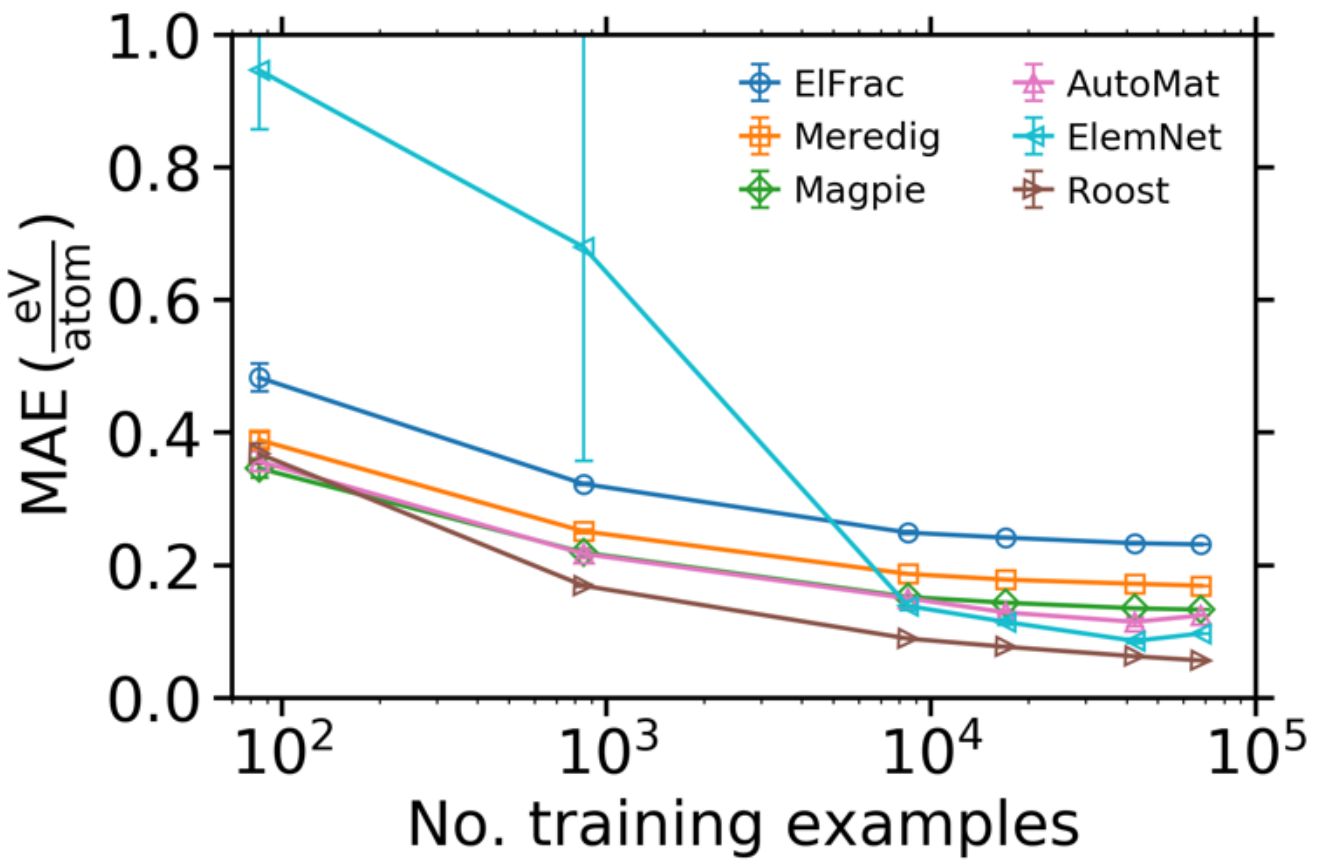

**Figure S5.** Learning curves for all compositional models. The MAE on predicting $\Delta H_f$ as a function of number of compounds used for training. Performance is shown on the test set, which is all MP compounds except those used for training. The MAE is averaged over five random splits of the training/testing compounds with the standard deviation in MAE over these five splits shown as the error bar. The final data point for each model at 68,011 training examples was taken from the 5-fold cross validation shown in **Figure 2**.

**Table S1.** Reproducing **Table 1** but training on $\Delta H_d$ instead of $\Delta H_f$.

| | *ElFrac* | *Meredig* | *Magpie* | *AutoMat* | *ElemNet* | *Roost* |
|---|---|---|---|---|---|---|
| **candidate compounds** | 13,659 | 13,659 | 13,659 | 13,659 | 13,659 | 13,659 |
| **stable compounds in MP** | 9 | 9 | 9 | 9 | 9 | 9 |
| **compounds predicted stable** | 0 | 0 | 0 | 0 | 58 | 299 |
| **% predicted stable** | 0 | 0 | 0 | 0 | 0.4 | 2.2 |
| **pred. stable and stable in MP** | 0 | 0 | 0 | 0 | 0 | 0 |

**Table S2.** The performance of each compositional representation trained to classify compounds as stable ($\Delta H_d \leq 0$) or unstable ($\Delta H_d > 0$). Note that the *Roost* representation is excluded from this analysis as described in Methods.

| | *Accuracy* | *$F_1$ score* | *False positive rate* |
|---|---|---|---|
| ***ElFrac*** | 0.723 | 0.631 | 0.191 |
| ***Meredig*** | 0.745 | 0.666 | 0.180 |
| ***Magpie*** | 0.759 | 0.683 | 0.170 |
| ***AutoMat*** | 0.792 | 0.732 | 0.153 |
| ***ElemNet*** | 0.744 | 0.683 | 0.219 |

**Table S3.** Training and inference times for learning and predicting $\Delta H_f$. Training time is the time required to train the models on 80% of the MP dataset (68,011 compounds). Inference time is the time required to predict $\Delta H_f$ for the remaining 20% of the MP dataset (17,013 compounds). Note that for *AutoMat*, the training time is a user-specified input.

| | *Training time (h)* | *Inference time (s)* |
|---|---|---|
| ***ElFrac*** | 0.02 | 15 |
| ***Meredig*** | 0.06 | 15 |
| ***Magpie*** | 0.05 | 15 |
| ***AutoMat*** | 10.00 | 2719 |
| ***ElemNet*** | 2.35 | 8 |
| ***Roost*** | 3.47 | 38 |
| ***CGCNN*** | 20.90 | 926 |